\documentclass[10pt,twocolumn,letterpaper]{article}

\usepackage[pagenumbers]{cvpr} % To force page numbers, e.g. for an arXiv version

\usepackage{graphicx}
\usepackage{amsmath}
\usepackage{amssymb}
\usepackage{booktabs}
\usepackage{float}
\usepackage{placeins}
\usepackage{stfloats}
\usepackage{amsmath}
\usepackage{xcolor}
\usepackage{enumitem}

\usepackage[pagebackref,breaklinks,colorlinks]{hyperref}

\newcommand\myparagraph[1]{\vspace{4pt} \noindent \textbf{#1.}}

\usepackage[capitalize]{cleveref}
\crefname{section}{Sec.}{Secs.}
\Crefname{section}{Section}{Sections}
\Crefname{table}{Table}{Tables}
\crefname{table}{Tab.}{Tabs.}

\definecolor{foo}{HTML}{EEEEFF}

\def\cvprPaperID{*****} % *** Enter the CVPR Paper ID here
\def\confName{URBANACCESS'24}
\def\confYear{2024}

\begin{document}

\title{Towards Data-Informed Interventions: Opportunities and Challenges of Street-level Multimodal Sensing}

\author{
\textbf{Joao Rulff$^{1}$, Giancarlo Pereira$^{1}$, Marcos Lage$^{2}$, }\\
\textbf{Maryam Hosseini$^{3}$, Claudio Silva$^{1}$}\\
\fontsize{10pt}{10pt}\selectfont $^{1}$~New York University, $^{2}$~Universidade Federal Fluminense, $^{3}$~University of California, Berkeley
}

\twocolumn[{%
\renewcommand\twocolumn[1][]{#1}%
\maketitle
\begin{center}
    \centering
    \captionsetup{type=figure}
    \includegraphics[width=\linewidth]{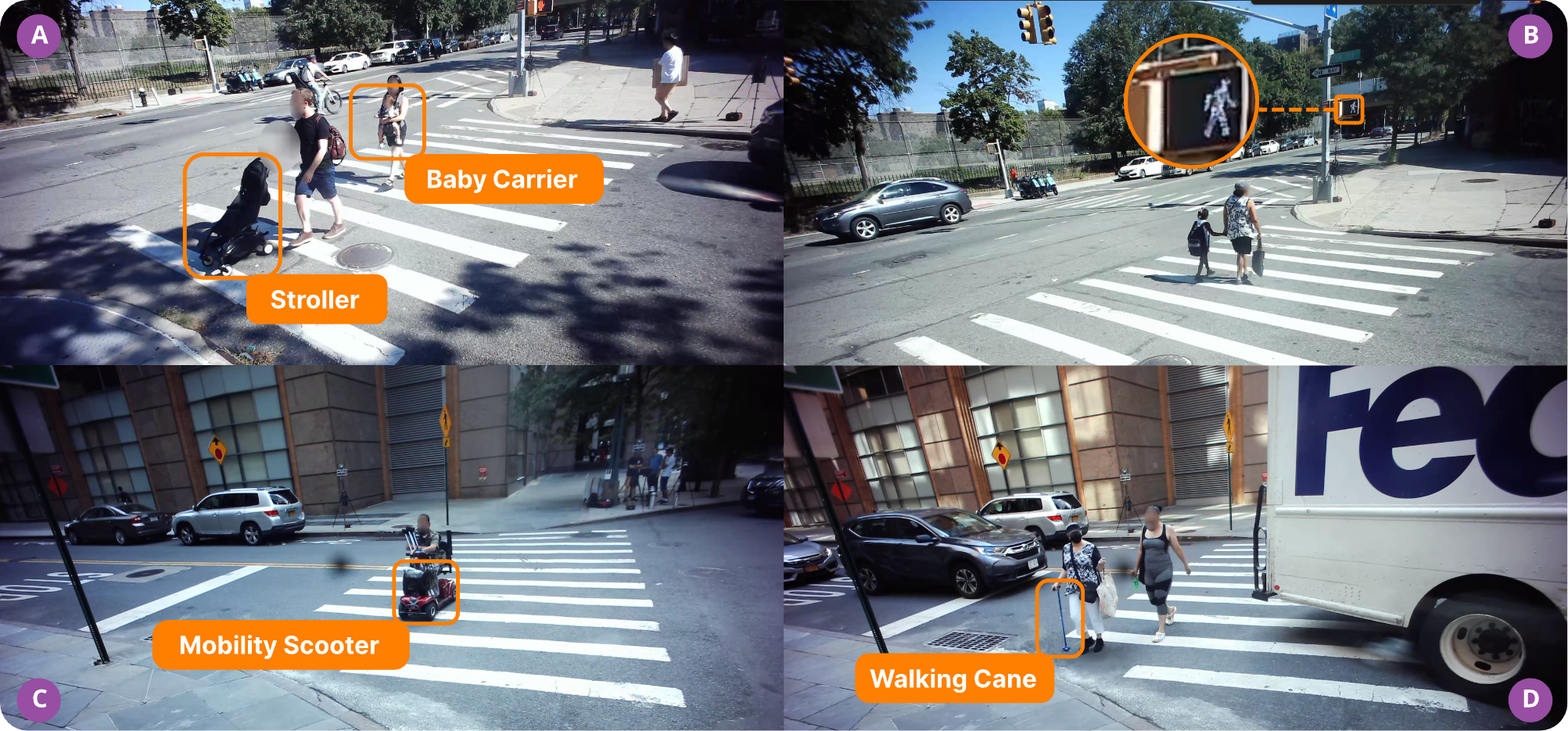}
    \captionof{figure}{This figure highlights situations where pedestrians with limited mobility levels cross streets (A, C, D). These examples were identified using open-vocabulary detection models. (B) shows how street-level videos can contain valuable information regarding traffic lights.}
    \label{fig:teaser}
\end{center}%
}]

%%%%%%%%% ABSTRACT
\begin{abstract}
Over the past decades, improvements in data collection hardware coupled with novel artificial intelligence algorithms have made it possible for researchers to understand urban environments at an unprecedented scale. From local interactions between actors to city-wide infrastructural problems, this new data-driven approach enables a more informed and trustworthy decision-making process aiming at transforming cities into safer and more equitable places for living.
This new moment unfolded new opportunities to understand various phenomena that directly impact how accessible cities are to heterogeneous populations. Specifically, sensing localized physical interactions among actors under different scenarios can drive substantial interventions in urban environments to make them safer for all. In this manuscript, we list opportunities and associated challenges to leverage street-level multimodal sensing data to empower domain experts in making more informed decisions and, ultimately, supporting a data-informed policymaking framework. The challenges presented here can motivate research in different areas, such as computer vision and human-computer interaction, to support cities in growing more sustainably. 
\end{abstract}

%%%%%%%%% MAIN BODY

\section{Introduction}\label{sec:intro}

In the complex system of urban public spaces, diverse actors, i.e., pedestrians, cyclists, and drivers, interact in dynamic ways.
Understanding these interactions has long been a goal of various fields ranging from transportation to social sciences. Computer vision methods opened new horizons for fine-level understanding of these dynamics by allowing scalable analysis of multimedia datasets.
However, accessibility-related challenges have often been overlooked. 
Assistive devices such as wheelchairs, walkers, and canes, which are crucial to many urban dwellers, have not been well represented in standard benchmarks like semantic segmentation or object detection. This oversight leaves a significant gap in how urban environments are understood and designed for inclusivity.
Recent developments, however, offer promising new avenues to address these gaps. 
Open-vocabulary object detection (OVD) models are emerging as a powerful tool to enhance inclusivity in computer vision tasks. Unlike traditional models limited by predefined categories, OVDs have the potential to detect and describe a far broader range of objects, including assistive devices. As represented in Figure~\ref{fig:teaser}, these models can quickly identify a set of cases where low-mobility pedestrians are interacting with urban intersections.

% Although these models represent a promising direction, their ability to reliably detect and classify assistive devices in diverse urban scenes remains an open question, necessitating further exploration.

In tandem, high-performance models for action recognition and pose estimation now allow researchers to track the behaviors and interactions of pedestrians with remarkable detail. These models, capable of identifying actions like walking, sitting, and running, offer critical insights into how pedestrians navigate the urban landscape, especially in scenarios where they interact with other actors, such as cyclists or vehicles. Multimodal data, including machine listening models, further enrich our understanding of urban dynamics by providing complementary information such as sound cues. Acoustic localization algorithms, which leverage microphone arrays, enable precise spatial mapping of sound events, enhancing the contextual understanding of city spaces.

Together, these advancements in both machine learning algorithms and hardware technologies are opening new opportunities to capture the fine-grained interactions that occur in urban environments. We are now able to leverage large datasets representing interaction among urban actors under different scenarios and analyze common patterns of specific populations, such as users of assistive tools, to propose interventions that aim to make these spaces safer and more accessible for underrepresented groups. These developments, however, also bring challenges that will likely drive research in areas such as computer vision and human-computer interaction in the coming years.

Building on our work deploying multimodal street-level sensors for large-scale data collection, we explore how the combination of audio and video data collected at the street level can offer deeper insights into urban accessibility and human behavior. In the sections that follow, we identify the key opportunities and challenges ahead, emphasizing the unique affordances that street-level multimodal datasets provide over traditional approaches.
\section{Related Work}\label{sec:rel}

Several efforts aim to collect multimedia urban data and use this data to reassess and build safer, more equitable, and more accessible urban spaces. Different research communities have studied and analyzed urban environments for decades using various approaches. However, large-scale urban multimedia datasets have only recently been made publicly available. 
Focusing on sound, SONYC \cite{SONYC} allows for a city-wide longitudinal understanding of human activity based on audio events. This dataset motivated several custom-made tools for efficiently exploring such modality, like Time Lattice \cite{Miranda_TimeLattice} and Urban Rhapsody \cite{Rulff2022}. Moving forward, initiatives like StreetAware \cite{StreetAware} and AIWaysion\footnote{https://www.aiwaysion.com} combine several modalities, including video, audio, and LiDAR, to support experts interested in understanding fine-grained dynamics of pedestrian activity, including crossing patterns. It is in the context of the experience of building the StreetAware dataset that this piece reports opportunities and challenges to leverage video and audio data to explore urban dynamics.
\section{Opportunities}\label{sec:oppo}

In this section, we elaborate on the opportunities we envision for extracting useful information from recent audiovisual datasets, such as StreetAware. We start by indicating its advantages over other types of data and then show how it can be used to improve pedestrian safety under various scenarios.

\subsection{Affordances of Audio and Video}

\myparagraph{Aerial versus Street-Level} As previous works present \cite{rs14030620}, aerial imagery can greatly support vehicular flow analysis. However, given its distance from objects of interest, we miss granular details that can enhance our comprehension of the scene. Pedestrian movement patterns, gait information, sidewalk materials, and street-level signal information represented in traffic lights are missed in this kind of data but are present in street-level imagery. This information is crucial to identify scenarios where pedestrians with disabilities are using different urban spaces. Enabling the automatic identification of classes related to pedestrians with disabilities (e.g., wheelchair and walker users) can support researchers in summarizing cases where these citizens are present and, therefore, understanding the common characteristics of these actors while interacting with various urban regions.

\myparagraph{LiDAR versus RGB Video}
LiDAR is a powerful technology that users laser imaging to quickly record data and convert build three-dimensional point clouds. This technology, however, still presents weaknesses for the work we want to accomplish in urban environments.

The dataset from Toronto-3D~\cite{tan2020toronto3d} is an example of comprehensive LiDAR data collection; it covers one kilometer of a road in Toronto, Canada and includes a dense point cloud with manually labelled segmentation of interesting features, such as road, building,and utility line.

The authors of Toronto-3D, however, have pointed out to know issues of the dataset involving moving cars.
For our research purposes, the motion of vehicles is crucial to explore the interaction between motorized and non-motorized actors in cities. 
With multiview RGB video data, we can track moving vehicles and easily compute direction of motion, speed, and how close these vehicles are to other agents.
These metrics are necessary for accurate time-to-collision calculations, which often serve as proxies to dangerous urban encounters.

\myparagraph{Audio}
Video and LIDAR data can be key to reconstructing digital representations of urban environments.
They are, however, limited to reconstructing what is in their field-of-view.
With audio data and audio detection models (such as YAMNet \cite{gemmeke2017audio}), we can "expand" the field-of-view of these sensors.
Events like an emergency vehicle approaching an intersection at high speed can benefit from audio detection, as the model would point out to these out-of-sight events to enrich data analyses that would otherwise be lost to limited field-of-view video data. 
This modality enables analysis aiming to understand the reaction time of different groups to out-of-sight events, which can support urban planners in creating tools to increase spatial awareness of groups of pedestrians with lower mobility levels.
Furthermore, audio localization techniques can support the visualization of spatial locations where drivers more often need to use attention-grabbing strategies, such as car honks. These can highlight spaces where interventions, like reducing the speed limit or improving signaling, can reduce the probability of accidents.  

\subsection{Movement Analysis}

Personal smart devices, such as smartwatches and smartphones, often measure people's movements throughout the day, reporting health metrics relevant to a person's well-being.
These data range from number of daily steps averaged in a week to heart rate throughout sleep. 
The data, collected by personal devices, are private and belong to the individuals themselves (in section~\ref{subsec:privacy} we discuss challenges surrounding privacy).

\begin{figure}[h]
  \centering
  \includegraphics[width=\linewidth]{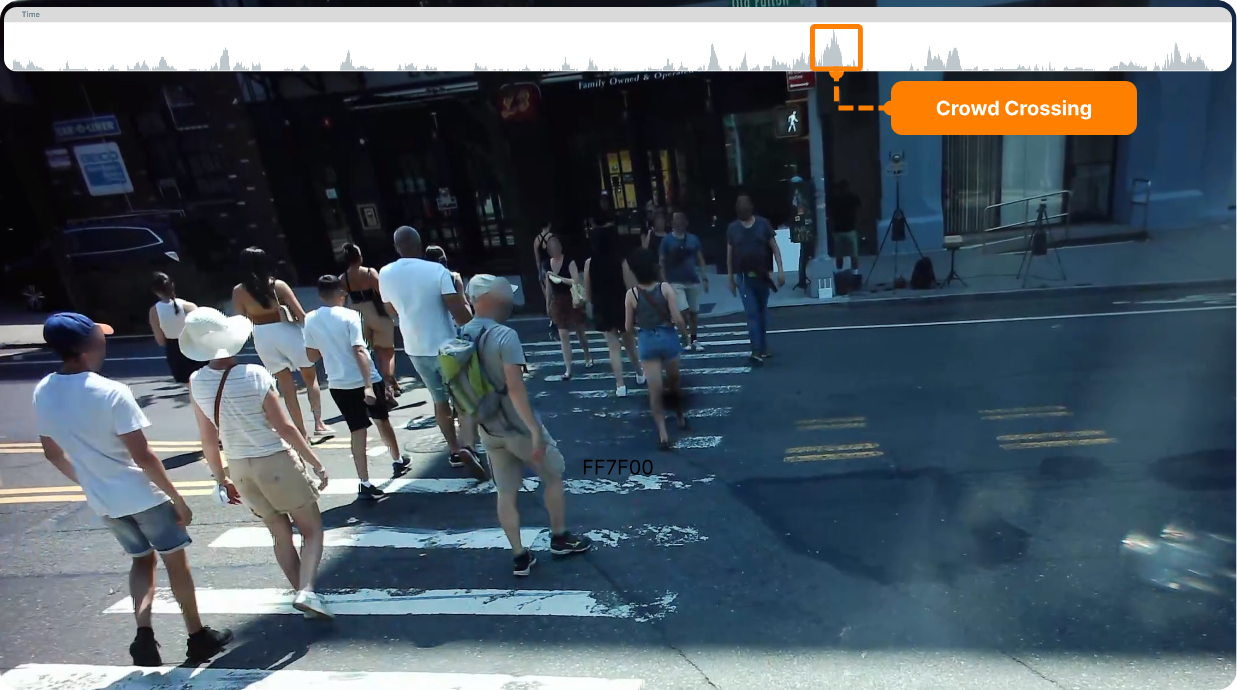}
  \caption{Distribution of pedestrian density over an entire video. Crowds are known to reduce the speed of pedestrians crossing the streets. Adapting traffic light timing based on pedestrian speed can be achieved by leveraging street-level RGB videos.}
\end{figure}

To extract similar movement data at large-scale in urban environments, we forego such personal devices and deploy our own cheap, multimodal sensors.
Advancements in deep learning this past decade now allow for human pose prediction models  \cite{Yulitehrnet21, hrnet_cheng2020, SunXLW19,Jocher_Ultralytics_YOLO_2023}.
With our RGB video data, we use such models to extract large-scale collections of two-dimensional poses of people on the streets.
Using multiple of our multimodal sensors recording from different views, we can then reconstruct people and capture their movements in the three-dimensional space.
We then extract metrics of people's movements patterns, such as speed of movement, direction, and gait.
For gait, for instance, we can measure distance between consecutive steps and the double support (DS) time.
DS time is the phase of the gait cycle where both feet are on the ground, allowing for greater stability and control of direction \cite{gait2021}.
Increased DS and large DS variability have both been associated with higher fear of falls and risks of falling \cite{Williams2019, HAUSDORFF20011050}. 
Therefore, we find that measuring gait of people accurately and at large scale in urban environments allows for urban planners to prepare and adapt the current infrastructure to benefit people of all motor skills.

This type of granular gait data also opens the opportunity to explore how people move in different weathers, obstacles, and pavement materials.
Project Sidewalk~\cite{projsidewalk2019} has developed a crowd-sourced, scalable data collection of sidewalk accessibility information using street-view images.
CitySurfaces~\cite{Hosseini2022} automatically detects, computes, and segments different pavement materials of sidewalk from street-level images.
Our work can complement these two works with the extraction of granular gait pattern of pedestrians. 
By combining the pavement materials, accessibility issues of sidewalks, and the gait data collected by our multimodal, street-level sensors, experts can now study at large-scales the movement pattern of individuals with distinct mobility levels on, for instance, sidewalks with surface problems during rainstorms.
These opportunities of data analyses highlight the role of urban planners and experts to better prepare our cities to accommodate different movement abilities in diverse (often adverse) environmental conditions and sidewalks with a range of accessibility issues.

\subsection{Adaptive Signal Timing}
Our proposal can open many pathways to interesting future developments.
With the advancement of Artificial Intelligence and faster network connectivity, street-level multimodal sensors can be integrated into the traffic system.

\myparagraph{Audio} Detecting audio events allows for a more dynamic and responsive traffic management, where, for instance, if the sensor hears an emergency vehicle siren approaching the intersection, it can safely adapt traffic and pedestrian crossings to allow for faster response time and reduce the chance of any unnecessary harm at the intersection. 

\myparagraph{RGB Video Data} With high-resolution cameras, we can monitor traffic light signals by extracting accurate traffic light cycle timing information. This new feature permits in-depth analysis of traffic management systems, crossing patterns, and vehicular flow at intersections. Street-level multimodal sensors can, in the future, be integrated with traffic management systems to allow pedestrians, especially those with lower mobility, to wait less and to cross the intersection with enough time. Using the aforementioned techniques to reconstruct gait patterns and to identify assistive tools correlated with lower mobility levels, such an intelligent system could leverage this information as a proxy for the presence of low-mobility actors to adapt traffic light times. 

% \subsection{Out-of-Sight Information}
% \textcolor{red}{delete?}
% The sensors can assist self-driving cars by sending information about events at the intersection \textcolor{red}{(like what?)}.

% Good 5G network required.

\section{Challenges}\label{sec:chal}

\subsection{Expert in the Loop}

Making these data useful for a diverse and non-technical community of domain experts is not trivial. Allowing these experts to directly interact with the data is an important step towards empowering them to make more informed decisions. 
However, extracting meaningful information from a mass of complex and large data is a difficult task. Therefore, designing intuitive visual analytics systems able to summarize and underline important details contained in the data is a must. 
These systems should, among other things, support cross-modal and interactive queries, requiring specialized data management infrastructure. Furthermore, tailored visualizations to enable the interpretation of different streams containing audio, visual, and derived metrics should be adequately tested against the target audience to assess their usability. 

\subsection{Optimal Deployment}
We have, for now, a finite number of sensors which can record one hour uninterruptedly.
This poses a significant bottleneck in our data acquisition.
We have continuously worked to improve these sensors, including higher camera resolution, longer battery life, and more resilient to extreme weather.
In order to obtain meaningful metrics of population with different abilities, the sensors need to be carefully placed around the city.
We expect that by selecting neighborhoods with distinct socioeconomic conditions and selecting locations near schools, hospitals, and assisted living houses, our studies will more likely represent people with different crossing abilities.
We also expect to highlight inequities in people's movement patterns around the city.
Some of these might include longer crossing wait times for pedestrians, more honks and higher decibel levels of noise, and poorer infrastructure and accessibility (such as lack of appropriate signage and crosswalk paint) for underresourced communities.
We also will guide the deployment by the occurrence of traffic violations and accidents around the city.
We aim to use publicly available data from \textit{NYC Open Data} to identify such landmarks and study which areas have a greater risk of harm that should be prioritized for sensor deployment. 

\subsection{Privacy}\label{subsec:privacy}
Our work entails recording hours of human behavior in urban environments.
Often, the metrics we are interested in exploring involve outliers and anomalies.
These can range from pedestrians crossing while distracted by their cellphones to cars violating traffic signaling.
We have taken the steps of anonymizing the video data by blurring faces and anonymizing the audio data by deleting recognizable speech.
This might prove insufficient with the general population, as they can rightfully be skeptical of such data collection in public spaces and might behave differently due to the presence of sensors in their surroundings.
Additionally, further study is needed to understand whether two-dimensional and three-dimensional poses can provide personally identifiable information.
Thus, implementing a citizen science agenda to better understand the general population's perception of privacy-preserving concerns will be useful in guiding future deployments and communication strategies.
\section{Conclusion}\label{sec:conc}

In this paper, we presented a set of opportunities the democratization of large-scale street-level multimodal data will enable over the next few years. Together with these, a set of challenges needs to be addressed to efficiently and responsibly support domain experts and policymakers in developing equitable and accessible regulations to support the safety and inclusion of every individual in the urban environment.

%%%%%%%%% REFERENCES
{\small
\bibliographystyle{unsrt}
\bibliography{references}
}

\end{document}